\documentclass[graybox]{svmult}

\usepackage{mathptmx}       
\usepackage{helvet}         
\usepackage{courier}        
\usepackage{type1cm}        
\usepackage{makeidx}         
\usepackage{graphicx}        
\usepackage{multicol}        
\usepackage[bottom]{footmisc}

\makeindex             

\begin{document}

\title*{Quadratic Palatini gravity and stable black hole remnants}
\author{Francisco S. N. Lobo, Gonzalo J. Olmo and D. Rubiera-Garcia}
\institute{F. S. N. Lobo \at Centro de Astronomia e Astrof\'{\i}sica da
Universidade de Lisboa, Campo Grande, Ed. C8 1749-016 Lisboa,
Portugal, \email{flobo@cii.fc.ul.pt}
\and G. J. Olmo \at Departamento de F\'{i}sica Te\'{o}rica and IFIC, Centro Mixto Universidad de
Valencia - CSIC. Universidad de Valencia, Burjassot-46100, Valencia, Spain \email{gonzalo.olmo@csic.es}
\and D. Rubiera-Garcia \at Departamento de F\'isica, Universidade Federal da
Para\'\i ba, 58051-900 Jo\~ao Pessoa, Para\'\i ba, Brazil \email{drubiera@fisica.ufpb.br}
}
%
%
\maketitle

\abstract{We present a four-dimensional Planck-scale corrected quadratic extension of General Relativity (GR) where no a priori relation between metric and connection is imposed (Palatini formalism). Static spherically symmetric electrovacuum solutions are obtained in exact analytical form. The macroscopic properties of these solutions are in excellent agreement with GR, though the region around the central singularity is modified. In fact, the singularity is generically replaced by a wormhole supported by the electric field, which provides a non-trivial topology to the space-time. Moreover, for a certain charge-to-mass ratio the geometry is completely regular everywhere. For such regular solutions, the horizon disappears in the microscopic regime below a critical number of charges ($N<17$), yielding a set of massive objects that could be naturally identified as black hole remnants.}

\section{Introduction}
\label{sec:1}

Almost one hundred years after its foundation, General Relativity (GR) has successfully passed a wide range of experimental tests \cite{Will:2005va}. However, there exist arguments suggesting that GR should be extended to encompass for new phenomena. One of them is represented by the existence of singularities both in the Big Bang and deep inside black holes, where the classical geometry breaks down. Such singularities are believed to be an artifact of the classical description, to be resolved within a quantum theory of gravity. Even though the specific details of such theory are not at hand, we could gain useful insights on its nature by studying classical gravity theories including higher-order curvature terms. Indeed such terms are needed for a high-energy completion of the Einstein-Hilbert Lagrangian \cite{field-quantization}, and moreover appear in some approaches to quantum gravity such as those based on string theory \cite{strings}. In this sense, if the underlying structure of spacetime is assumed to be Riemmanian \emph{a priori} such terms usually give rise to fourth-order field equations and bring in ghosts. However, the fact is that metric and connection are independent entities by their own, and thus, in absence of further evidence for the physics beyond GR, one could apply the variational principle to both objects, instead of relaying on a tradition or convention basis to keep the Riemannian assumption at all energies/scales. In this (Palatini) formulation of the theory, quadratic gravity Lagrangians satisfy second-order equations and contain Minkowski as a vacuum solution of the theory, which guarantees the absence of ghosts.

\section{Theory and solutions}
\label{sec:2}

We thus consider Palatini Lagrangian densities defined as
\begin{equation}
S=\frac{1}{2\kappa^2} \int d^4 x \sqrt{-g} f(R,Q) + S_M[g_{\mu\nu}, \Psi] \label{eq:action}
\end{equation}
with $R=g_{\mu\nu}R^{\mu\nu}$, $Q=R_{\mu\nu}R^{\mu\nu}$ and $S_m$ the matter action. We apply the variation principle independently to metric and connection, and impose vanishing torsion $\Gamma_{[\mu\nu]}^{\lambda}=0$ {a posteriori}, which leads to $R_{\mu\nu}=0$ and to the existence of volume invariants preserved by the theory \cite{or13}. Under these conditions the field equations become

\begin{eqnarray}
\delta g_{\mu\nu} \rightarrow f_R R_{\mu\nu}-\frac{f}{2}g_{\mu\nu}+2f_QR_{\mu\alpha}{R^\alpha}_\nu &=& \kappa^2 T_{\mu\nu} \label{eq:metric} \\
\delta  \Gamma_{\mu\nu}^{\lambda} \rightarrow \nabla_{\beta}\left[\sqrt{-g}\left(f_R g^{\mu\nu}+2f_Q R^{\mu\nu}\right)\right]&=&0  \label{eq:connection}
\end{eqnarray}
To solve them we note that the equation for the connection allows to define a rank-two tensor $h_{\mu\nu}$ as $\nabla_{\beta} [\sqrt{-h} h^{\mu\nu}]=0$, related to the metric $g_{\mu\nu}$ as

\begin{equation}
h^{\mu\nu}=\frac{g^{\mu\alpha} {\Sigma_\alpha}^{\nu}}{\sqrt{det \Sigma}} \hspace{0.1cm}; \hspace{0.1cm} h_{\mu\nu}= (\sqrt{det \Sigma}) [\Sigma^{-1}]_{\mu}^{\alpha} g_{\alpha \nu} \hspace{0.1cm}; \hspace{0.1cm} {\Sigma_{\alpha}^{\nu}}=(f_R \delta_{\alpha}^{\nu} + 2f_Q {P_\alpha}^{\nu}) \label{eq:h-to-g}
\end{equation}
and thus the independent connection $\Gamma_{\mu\nu}^{\lambda}$ is the Levi-Civita connection of $h_{\mu\nu}$. In terms of $h_{\mu\nu}$ the equation for the metric (\ref{eq:metric}) become \cite{or12a}

\begin{equation}
{R_\mu}^{\nu}(h)=\frac{1}{\sqrt{det \Sigma}} \left(\frac{f}{2} \delta_{\mu}^{\nu} + \kappa^2 {T_\mu}^{\nu} \right) \label{eq:eom}
\end{equation}
This shows that the \emph{auxiliary} metric $h_{\mu\nu}$ satisfies an Einstein-like set of partial differential equations. Moreover, since $g_{\mu\nu}$ is algebraically related with $h_{\mu\nu}$, the field equations for $g_{\mu\nu}$ are also second-order. Note also that in vacuum ${R_\mu}^{\nu}=\Lambda \delta_{\mu}^{\nu}$ and thus we recover GR with possibly a cosmological constant term. This shows that these theories have the same number of propagating degrees of freedom as GR.

Next we consider the quadratic model $f(R,Q)=R+l_P^2(R+aQ)$, with $l_P^2\equiv\hbar G/c^3$ representing the Planck length squared and $a$ is a free parameter, and to probe the theory we study electrovacuum solutions, as given by the action

\begin{equation}
S_m[g,\psi_m] = -\frac{1}{2\kappa^2}\int d^4x \sqrt{-g}F_{\alpha\beta}F^{\alpha\beta}
\end{equation}
where $F_{\mu\nu}=\partial_{\mu}A_{\nu}-\partial_{\nu}A_{\mu}$ is the field strength tensor. The energy-momentum tensor reads $ {T_\mu}^\nu=-\frac{1}{4\pi}[{F_\mu}^\alpha {F_\alpha}^\nu-\frac{{F_\alpha}^\beta{F_\beta}^\alpha}{4}\delta_\mu^\nu]$, while the field equations $\nabla_{\mu}(F^{\mu\nu})=0$ in a spherically symmetric background $ds^2=g_{tt}dt^2+g_{rr}dr^2+r^2d\Omega^2$ satisfy $F^{tr}=\frac{q}{r^2}\frac{1}{\sqrt{-g_{tt}g_{rr}}}$. For this theory, the field equations (\ref{eq:eom}) in the geometry $h_{\mu\nu}$ can be analytically solved, and putting that solution in terms of $g_{\mu\nu}$ using (\ref{eq:h-to-g}) leads to \cite{or12a,or12b,or12c}

\begin{equation}
g_{tt}=-\frac{A(z)}{\sigma_+} \ , \ g_{rr}=\frac{\sigma_+}{\sigma_-A(z)}  \ , \ A(z)=1-\frac{\left[1+\delta_1 G(z)\right]}{\delta_2 z \sigma_-^{1/2}} \ , \ \frac{dG}{dz}=\frac{z^4+1}{z^4\sqrt{z^4-1}} \nonumber
\end{equation}
where $z=r/r_c$, $r_c\equiv \sqrt{r_q l_P}$, $r_q^2\equiv \tilde{\kappa}^2q^2$, $\delta_1=\frac{1}{2r_S}\sqrt{r_q^3/l_P}, \delta_2= \sqrt{r_q l_P}/r_S$ and $\sigma_{\pm}=1 \pm 1/z^4$. The function $G(z)$  behaves as follows: far from the center we have $G(z)\approx -\frac{1}{z}-\frac{3}{10z^5}+\ldots$ and thus the GR limit is quickly recovered for $z\gg 1$. However, at $z\to 1$ the metric undergoes important modifications, as here we obtain $G(z)\approx \beta+2 \sqrt{z-1}-\frac{11}{6} (z-1)^{3/2}+\ldots $ where $\beta\approx -1.74804$ is a constant needed to match the far and near expansions. At $z=1$, in general, the curvature invariants become divergent, but for the particular choice $\delta_1=\delta_1^*=-1/\beta$ they behave as

\begin{eqnarray}
r_c^2 R(g)\approx  \left(-4+\frac{16 \delta _c}{3 \delta _2}\right)+\ldots \hspace{0.1cm} &;& \hspace{0.1cm}
r_c^4 Q(g) \approx   \left(10+\frac{86 \delta _c^{2}}{9 \delta _2^2}-\frac{52 \delta _c}{3 \delta _2}\right)+\ldots \nonumber \\
r_c^4K(g)& \approx &  \left(16+\frac{88 \delta _c^{2}}{9 \delta _2^2}-\frac{64 \delta _c}{3 \delta _2}\right)+\ldots \nonumber
\end{eqnarray}
and thus all of them are finite. Thus, charged solutions in the quadratic Palatini theory (\ref{eq:action}) somehow interpolate between the Schwarzschild ($\delta_1<\delta_c$) and Reissner-Nordstr\"om ($\delta_1>\delta_c$) solutions of GR. However, the non-singular solutions with $\delta_1=\delta_1^*$ are genuine predictions beyond GR.

\section{Wormhole structure and black hole remnants}

Using Eddington-Finkelstein type coordinates, $ds^2=g_{tt} dv^2+2dv dr^*+r^2(r^*)d\Omega^2$, the line element at the surface $z=1$ becomes

\begin{equation}
g_{tt}= \frac{\left(1-\delta _1/\delta_c\right)}{4\delta _2 \sqrt{z-1}}-\frac{1}{2}\left(1-\frac{\delta _1}{ \delta _2}\right)+O(\sqrt{z-1})
\end{equation}
For $\delta_1=\delta_c$, we find that $g_{tt}$ is smooth everywhere, and $r^*$ can be extended to the negative axis by considering the two branches of the equation $(dz^*/dz)^2= 1/\sigma_{-}$, where $\sigma_-$ gives the relation between the $2$-spheres of $g_{\mu\nu}$ and $h_{\mu\nu}$. When $\delta_1=\delta_c$ the radial function has a minimum at $r=r_c$ (or $z=1$), and thus the bounce in the coordinate $z^*$ is achieved through $z^*={_{2}F}_1[-\frac{1}{4},\frac{1}{2},\frac{3}{4},\frac{1}{z^4}] z$ if $z^*\ge z^*_c$ and $z^*=2z^*_c-{_{2}F}_1[-\frac{1}{4},\frac{1}{2},\frac{3}{4},\frac{1}{z^4}] z$ if $z^*\le z^*_c $. This is reminiscent of a wormhole geometry, with $z=z_c$ the wormhole throat. This wormhole structure naturally leads to the geon concept, introduced by Wheeler in 1955 as ``sourceless self-consistent gravitational electromagnetic entities" with a multiply-connected geometry \cite{Wheeler1955}. This implies that the lines of force enter through one of the mouths of the wormhole and exits through the other, creating the illusion of a negatively charged object on one side and a negatively charged on the other. The locally measured electric charge comes from the computation of the flux $\Phi\equiv \int_S *F=4\pi q$ through any $2-$surface $S$ enclosing the wormhole throat.  This implies that the electric flux per surface unit at $z=1$, which represents the density of lines of force crossing the wormhole throat, namely, $\Phi/(4\pi r_c^2)=q/r_c^2=\sqrt{c^7/(2(\hbar G)^2)}$, turns out to be an \emph{universal quantity}, independent of the specific amounts of charge and mass. Moreover, the dependence of $\frac{\Phi}{4\pi r_c^2}$ on fundamental constants (irrespective of $M$ and $q$) indicates that the geon structure persists for those solutions showing curvature divergences.

The evaluation of the action on the solutions yields

\begin{equation}
S_T=\frac{q^2}{r_c \delta_1^*}\int dt=2\frac{\delta_1}{\delta_1^*}M c^2 \int dt
\end{equation}
which can be interpreted as the addition of the electromagnetic plus {\it gravitational binding energy}, for the $\delta_1=\delta_c$  solutions. This finite value coincides with the action of an isolated particle of mass $M$ at rest: $S_{p.p.}=mc^2\int dt\sqrt{1-v^2/c^2}$.

One might argue that the event horizon is expected to force the decay into $\delta_1\neq \delta_c$ states. However, at $z=1$, the behaviour of
\begin{equation}
g_{tt}= \frac{\left(1-\delta _1/\delta_c\right)}{4\delta _2 \sqrt{z-1}}-\frac{1}{2}\left(1-\frac{\delta _1}{ \delta _2}\right)+O(\sqrt{z-1}) \label{eq:solution}
\end{equation}
implies that the sign of $\left(1-\delta _1/\delta_c\right)$ determines if there is an event horizon. Expressing $q=N_q e$, we find that  $r_q=2l_P N_q/N_q^c$, where $N_q^c\equiv\sqrt{2/\alpha_{em}}\approx 16.55$. Thus, if $\delta_c/\delta_2=N_q/N_q^c>1$ there is an event horizon, which is absent otherwise. In summary, for $\delta_1= \delta_c$  we find a family of completely stable objects with mass spectrum: $M\approx 1.23605 (N_q/N_q^c)^{3/2} m_P$.  The absence of an event horizon for $N_q<16.55$ yields quantum mechanically stable objects. The topological nature of their charge makes them stable against arbitrary classical perturbations that preserve the topology.

It is worth mentioning that Hawking's estimates \cite{H1}, based entirely on the process of classical collapse, are in excellent quantitative agreement with our results. Thus, large numbers of objects with $M\sim m_P$ and $N_q < 30 $ could have been formed in the early universe, and a fraction of them could reach the stability conditions found here.  The evaporation of more massive objects could also yield this type of particles. The existence of stable states in the lowest part of the mass and charge spectrum which can be continuously connected with black hole states, supports the view that these objects can be naturally identified as black hole remnants. The existence of these objects suggests that a maximum temperature should be reached in the evaporation process, which is compatible with the lack of observations of black hole explosions. Note also that the robustness of our results can be tested by considering nonlinearities (coming e.g. from pair production) in the matter sector through nonlinear electrodynamics, finding that exact solutions, provided by Born-Infeld electrodynamics, are in qualitative agreement with the Maxwell case \cite{or13b}. Moreover, it turns out that the mass spectrum can be reduced by many orders of magnitude.

\section{Dynamical generation}
\label{sec:3}

The above wormhole solutions can be dynamically generated as follows. Consider a presureless flux of ingoing charged matter with stress-energy tensor
$T_{\mu\nu}^{flux}= \rho_{in} k_\mu k_\nu$. This flux conveys a current $J^\nu\equiv \Omega(v) k^\nu$ (with $\Omega(v)$ a function to be determined) in the electromagnetic equations of motion as $\nabla_\mu F^{\mu\nu}=4\pi J^\nu$. The field equations (\ref{eq:eom}) admit a solution \cite{lor13b}, which is formally similar to that of Eq.(\ref{eq:solution}) with the replacements $t \rightarrow v$, $\delta_1 \rightarrow \delta_1(v)$, $\delta_2 \rightarrow \delta_2(v)$, and $g_{tt} \rightarrow g_{tt} + 2l_P^2 \kappa^2 \rho_{in}/(\sigma_{-}(1-2r_c^4/r^4))$, where $v$ is the advanced time coordinate.

If we assume the initial state to be Minkowski space, a charged perturbation of compact support propagating within an interval $[v_i,v_f]$ makes the area of the $2$-spheres of constant $v$ to never become smaller than $r_c^2 (v)$. Once the flux is off ($v>v_f$), the result is a static geometry identical to that given by Eq.(\ref{eq:solution}). This change in the geometry can be interpreted as the formation of a wormhole whose throat has an area $A_{WH}=4\pi r_c^2(v_f)$.  The existence or not of curvature divergences at $r=r_c(v_f)$ depends on the (integrated) charge-to-mass ratio of the flux. This result seems to imply a change in the global properties of space-time and, in particular, in its topology. However, a different viewpoint emerges if one accepts that curvature divergences are not necessarily as pathologically as usually regarded. Indeed, the existence of a well defined electric flux flowing accross the wormhole throat support the idea that the wormhole structure is present event in those cases with curvature divergences. Within this interpretation, the region the region $v<v_i$ would be made of two disconnected pieces of Minkowski space-time which are separated by a vanishing radius throat that, once electric charge comes into play, is opened up.

\section{Ending comments}
\label{sec:4}

An important lesson that follows from our analysis is that given the puzzles that Nature is offering to us in the understanding the physics beyond GR, the consideration of foundational aspects of gravity, such as the validity of the Riemannian assumption on the spacetime at all scales/energies, cannot be overlooked any longer. This important aspect has consequences on the formulation of classical gravity theories, as the metric and Palatini approach yield, in general, inequivalent results \cite{Borunda} (an important exception is precisely GR). Working in the Palatini approach we have found the existence of wormhole solutions with a geonic structure that can be interpreted as black hole remnants. The existence of a non-trivial topology conveyed by these solutions raises several questions on the true meaning of curvature divergences in this theory, and on the very nature of space-time that requires further investigation.

Work supported by FCT grants CERN/FP/123615/2011 and CERN/FP/123618/2011 (F.S.L.), grants FIS2011-29813-C02-02, the
Consolider Programme CPAN (CSD2007-00042), and the JAE-doc program of CSIC (G.J.O.) and Brazilian grant 561069/2010-7 (D.R.-G.).

\end{document}